\documentclass[aps,twocolumn,nofootinbib]{revtex4}
\usepackage[latin1]{inputenc}
\usepackage{epsfig}
\newcommand{\beq}{\begin{equation}}
\newcommand{\eeq}{\end{equation}}
\newcommand{\la}{\langle}
\newcommand{\ra}{\rangle}

\begin{document}

\title{Irreversible thermodynamics and 
Glansdorff-Prigogine principle derived from
stochastic thermodynamics}

\author{Tânia Tomé and Mário J. de Oliveira}
\affiliation{Universidade de São Paulo,
Instituto de Física,
Rua do Matão, 1371, 05508-090
São Paulo, SP, Brasil}

\begin{abstract}

We derive the main equations of irreversible thermodynamic
including the expression for the Glansdorff-Prigogine extremal
principle from stochastic thermodynamics. To this end, we analyze
a system that is subject to gradients of temperature and external
forces that induce the appearance of fluxes of several sorts and
the creation of entropy. We show that the rate of entropy
production is a convex function of the fluxes, from which follows 
that the excess entropy production is nonnegative, which is an
expression of the Glansdorff-Prigogine principle. We show that
the Lyapunov function associated with the excess entropy production
can be identified with a thermodynamic potential in the special
case where the gradients of temperature are absent.

\end{abstract}

\maketitle

\section{Introduction}

Irreversible thermodynamics \cite{prigogine1947,
denbigh1951,prigogine1955,meixner1959,degroot1962,fitts1962,
glansdorff1971,nicolis1977,kondepudi1998,lebon2008}
is a macroscopic theory that deals with systems in states out
of thermodynamic equilibrium. These states are maintained by
gradients of temperature that induce heat and entropy fluxes
and by external forces that cause the appearance of other types
of fluxes.
The change in energy of the system is due to the heat
flux and by the power of external forces.
The change in entropy of the system is not only due
to the entropy fluxes but also due to the creation of entropy
caused by irreversible processes occurring inside the system.

A system out of equilibrium is characterized within irreversible
thermodynamics by variables that include the fluxes of various
sorts and also by those variables that define the state of
thermodynamic equilibrium such as the energy and entropy of
the system. These quantities varies in time and eventually
approach a final value in the stationary state. The time variation
of the energy of the system is equal to its flux into the system
because energy is a conserved quantity.
However, this is not the case of entropy, which may be
created.
The time variation of the entropy of the system is thus equal to
the rate of entropy production minus the entropy flux to
the outside. The fundamental property of the production
of entropy is that it is nonnegative, which is an
expression of the second law of thermodynamics.

In a state of thermodynamic equilibrium, there is no
production of entropy. A system out of thermodynamic
equilibrium is characterized by a continuous production
of entropy. When the system approaches a stationary state
the production of entropy reaches a value which, according
to the extremal principle introduced by Prigogine in 1945
\cite{prigogine1945}, is a minimum. He based the principle
on the linear relation between forces and fluxes and on
the Onsager reciprocity relations \cite{onsager1931}.

When the condition of linearity between forces and fluxes
are not valid, as happens if the system is not close to
equilibrium, one does not expect the principle to be valid.
This lead Glansdorff and Prigogine to formulate in 1954
\cite{glansdorff1954} a more general extremal principle,
expressed in the following terms 
\beq
\sum_k \delta X_k \,\delta J_k \geq 0,
\label{12}
\eeq
where $\delta J_k$ and $\delta X_k$ are, respectively,
the deviations of the fluxes $J_k$ and the forces $X_k$
from their values at the stationary state.

We aim here to derive the equations of irreversible
thermodynamics, including the Glandorff-Prigogine principle,
from stochastic thermodynamics\cite{tome2006,tome2010,
esposito2010,vandenbroeck2010,vandenbroeck2013,tome2015}.
Our main result concerns the convexity property of the
rate of entropy production. We show that this quantity is
an upward convex function of the collection of fluxes.
This property allows us to defined an excess entropy
production ${\cal P}_{\rm exc}$ which is also upward
convex and a minimum at the stationary state, which
is an equivalent statement of the Glansdorff-Prigogine
principle.

The Glansdorff-Prigogine principle has been understood as
a  criterion for the stability of the stationary state,
and in this sense it has been regarded as connected to a
Lyapunov function \cite{schlogl1971a,tomita1972,desobrino1975,
schnakenberg1976, jiuli1984,maes2015,ito2022}. This 
connection is expressed by the relation
\beq
\frac{d L}{dt} = {\cal P}_{\rm exc},
\eeq
where $L$ is the Lyapunov function understood as a function
of probabilities of the microstates
\cite{schlogl1971a,schnakenberg1976}. 
In this sense $L$ is not in general a function of the
macroscopic thermodynamic variables. However, as we will show here,
there is a special case, namely, when there is no temperature
gradients, that this is possible. In this case the Lyapunov
function is identified as a thermodynamic potential.

In the next section we formulate the irreversible thermodynamics
as a macroscopic theory and present its main results, including
the formulation of the Glansdorff-Prigogine principle in terms
of the excess entropy production. In the subsequent chapter we
demonstrate from stochastic thermodynamics the propositions
that were introduced as assumptions and postulates in the
present formulation of irreversible thermodynamics,
including the convexity of the entropy production.

\section{Irreversible thermodynamics}

\subsection{Fluxes and forces}
\label{CL}

A system out of thermodynamic equilibrium that is the object
of study of irreversible thermodynamics is described by the
fluxes of several sorts. In addition to the fluxes the
system is also described by those variables that define
the state of the system when in equilibrium. 
These include the entropy $S$, the internal energy $U$,
and a set of complementary macroscopic variables
$N_1,N_2,\ldots,N_c$. These variables vary in time as a
consequence of two classes of processes. One class consists
of the internal processes and the other consists of the
processes caused by the external forces, which
we call external processes.

The variation in time of $N_l$ has two contributions.
One is the flux $\chi_k$ from the outside induced by
the external process $l$ and the other is a
term $R_l$ describing its creation or annihilation 
due to {\it all} internal processes. Therefore,
\beq
\frac{dN_l}{dt} = \chi_l + R_l.
\label{25}
\eeq
Each external process $l$ ensues a work done on the system
per unit time, which is proportional to $\chi_l$, that is,
equal to $\mu_l \chi_l$ where $\mu_l$ is a parameter.
The total work done on the system per unit time by the
external forces is 
\beq
\Phi_{\rm w} = \sum_l \mu_l \chi_l,
\label{45}
\eeq
where the summation is over the external processes.

The time variation of the internal energy is also due to
the internal and external processes. We denote by $\phi_k$
the consumption of energy per unit time due to the process
$k$, either internal or external. Therefore, the time
derivative of the energy is given by
\beq
\frac{dU}{dt} = \Phi_{\rm u},
\label{39a}
\eeq
where 
\beq
\Phi_{\rm u} = \sum_k \phi_k,
\label{39b}
\eeq
and the summation is over all processes, internal or external.

The flux of heat $\Phi$ into the system is equal to the time
variation of the internal energy minus the work done on the system,
\beq
\Phi=\Phi_{\rm u}-\Phi_{\rm w},
\eeq
which is the expression of the first law, or
\beq
\Phi = \sum_k \phi_k + \sum_l (\phi_l-\mu_l \chi_l),
\label{27}
\eeq
where the first and the second summation run over
the internal and the external processes, respectively.

The entropy $S$ is not a conserved quantity, but it cannot be
annihilated. Thus its variation in time is equal to the rate
at which it is being created, denoted by ${\cal P}$, minus the
flux of entropy to the outside $\Psi$, which is  expressed by
\beq
\frac{dS}{dt} = {\cal P} - \Psi.
\label{29}
\eeq
The rate of entropy production is nonnegative,
\beq
{\cal P}\geq0,
\label{19}
\eeq
which is the expression of the second law and is
a {\it postulate} of irreversible thermodynamics.

The expression for the entropy flux $\Psi$ is set up as follows.
Each term of the first summation in (\ref{27}) is understood as
a part of the total heat flux $\Phi$ that is being introduced
into the system from a section of the environment that is
understood as a heat bath at a temperature $T_k$. This results
in a contribution to the entropy flux to the system which
we assume to be equal to $\phi_k/T_k$. 
In an analogous manner the contribution
to the entropy flux coming from the terms of the second
summation in (\ref{27}) is $(\phi_k-\mu_k \chi_k)/T_k$.
Therefore, the entropy flux is written as
\beq
\Psi = - \sum_k \frac1{T_k}\phi_k
- \sum_l \frac{1}{T_l}(\phi_l - \mu_l \chi_l),
\label{21}
\eeq
the minus sign being introduced because $\Psi$ is the flux 
{\rm from} the system {\it to} the outside. The first and
the second summation run over the internal and the external
processes, respectively.
The relation between entropy flux and entropy flux that we
have just assumed is a {\it postulate} of the 
present formulation of irreversible
thermodynamics that we call Clausius relation.

We write equation (\ref{21}) as
\beq
\Psi = - \sum_k \frac1{T_k}\phi_k
+ \sum_l \frac{\mu_l}{T_l} \chi_l,
\label{21a}
\eeq
where now the first summation runs over all processes. This
expression allows us to introduce the following simplification.
We denote by $J_k$ the negative of the fluxes, that is, $J_k$
can be either $-\phi_k$ or $-\chi_l$. The quantities that
multiply these quantities in the expression (\ref{21a}) are
the {\it thermodynamic forces}, denoted by $F_k$. That is,
$F_k$ can be either $1/T_k$ or $-\mu_l/T_l$. Using this
notation, we write the entropy flux given by equation
(\ref{21a}) as the bilinear form
\beq
\Psi = \sum_k F_\nu J_k.
\label{22}
\eeq

We remark that the thermodynamic forces $F_k$ are understood
as parameters and should not be confused with the forces $X_k$
appearing in the expression (\ref{12}). These quantities, that
we call {\it conjugate forces}, are defined by 
\beq
X_k = \frac{\partial{\cal P}}{\partial J_k},
\label{23}
\eeq
where ${\cal P}$ is understood
as a function of the fluxes $J_k$.

\subsection{Glansdorff-Prigogine principle}
\label{GP}

To reach expression (\ref{12}) of the Glansdorff-Prigogine
principle, we assume that the production of entropy
${\cal P}$ is an {\it upward convex function} of
the set of fluxes $J_k$. This assumption is also a
{\it postulate} of the present approach of irreversible
thermodynamics. Defining 
\beq
A_{kl} = \frac{\partial X_k}{\partial J_l} =
\frac{\partial^2{\cal P}}{\partial J_k\partial J_l},
\label{10}
\eeq
the convexity of ${\cal P}$ implies that the matrix
$A$ with elements $A_{kl}$ is semi-positive definite, 
which is equivalent to say that
\beq
\sum_{kl} A_{kl}\,\delta J_k\, \delta J_l \geq 0,
\label{14a}
\eeq
where $\delta J_k$ are deviations of the fluxes $J_k$.

Taking into account that the variation of $X_k$ is
\beq
\delta X_k = \sum_l A_{kl} \delta J_l,
\eeq
which follows from (\ref{23}), we reach the expression
\beq
\sum_k \delta X_k \delta J_k \geq 0.
\eeq
This is the expression (\ref{12}) of the Glansdorff-Prigogine
principle provided we interpret $\delta J_\mu$ as the deviations
of the flux from their stationary values.

The Glansdorff-Prigogine principle can be formulated in an
equivalent manner in terms of the excess entropy production
defined by
\beq
{\cal P}_{\rm exc} = {\cal P} - \sum_k X_k^0 (J_k - J_k^0)
- {\cal P}_0,
\label{28}
\eeq
where $X_k^0$, $J_k^0$, and ${\cal P}_0$, are respectively
the values of the conjugate forces, the fluxes, and the
entropy production at the stationary state. From the
definition (\ref{28}) it follows that the first order variation
$\delta {\cal P}_{\rm exc}$ from the stationary state
vanishes. Since ${\cal P}$ is upward convex function so
is ${\cal P}_{\rm exc}$ because it differs from ${\cal P}$
by linear terms in $J_k$, that is,
$\delta^2{\cal P}_{\rm exc} = \delta^2{\cal P} \geq 0$.
Taking into account that $\delta{\cal P}_{\rm exc}=0$,
then it follows that the excess entropy production is a
minimum at the stationary state, in fact an absolute minimum, 
which is an equivalent form of the Glansdorff-Prigogine
principle. We may write
\beq
{\cal P}_{\rm exc} \geq 0,
\eeq
because from its definition, the value of the excess entropy
production at the stationary state is zero.

\subsection{Thermodynamic potential}

Let us replace the expression for ${\cal P}$ coming from
(\ref{28}) and the expression for $\Psi$ given by (\ref{22})
in equation (\ref{29}). The result is
\beq
\frac{dS}{dt} + \sum_k (F_k - X_k^0)(J_k - J_k^0)
= {\cal P}_{\rm exc},
\label{34}
\eeq
where we have taken into account that 
\beq
{\cal P}_0 = \Psi_0 = \sum_k F_k J_k^0.
\eeq
The left-hand side of (\ref{34}) is not in general the time
derivative of a thermodynamic potential, which here we are
defining as any linear combination of $S$, $U$, and $N_l$.
However, this happens when all temperatures are equal,
as we show next.

When the temperatures $T_k$ and $T_l$ are all the same,
the expression (\ref{21}) for the entropy flux becomes 
\beq
\Psi = - \frac1T \phi_{\rm u}  + \frac1T \sum_k \mu_k \chi_k,
\eeq
where $T$ is the common temperature, 
and the excess entropy production given by (\ref{28}) becomes
\beq
{\cal P}_{\rm exc} = {\cal P}
- \sum_k X_k^0 (\chi_k - \chi_k^0)- Y^0 \Phi_{\rm u}
- {\cal P}_0,
\eeq
where $Y=-\partial{\cal P}/\partial \Phi_{\rm u}$
and $X_k=-\partial{\cal P}/\partial \chi_k$.
The equation (\ref{34}) becomes
\beq
\frac{dS}{dt} + \sum_k (\frac{\mu_k}{T} + X_k^0)\frac{dN_k}{dt}
- (\frac1T - Y^0)\frac{dU}{dt}
= {\cal P}_{\rm exc},
\eeq
and we see that the left-hand side is the time derivative 
of the thermodynamic potential
\beq
M = S + \sum_k (\frac{\mu_k}{T} + X_k^0) N_k
- (\frac1T - Y^0) U,
\eeq
that is,
\beq
\frac{dM}{dt} = {\cal P}_{\rm exc}.
\eeq   
Recalling that ${\cal P}_{\rm exc}\geq0$, we find
\beq
\frac{dM}{dt} \geq 0,
\eeq
and the thermodynamic potential $M$ increases with
time towards its value at the stationary state.

\section{Stochastic thermodynamics}

\subsection{Transition rates}

We consider the same system studied in the previous section
but now we use a microscopic description provided by the
stochastic thermodynamics. The evolution of the system
follows a stochastic dynamics in continuous time governed
by a master equation. The probability distribution $p_i$ of
microstates $i$ evolves in time according to the master equation
\beq
\frac{dp_i}{dt} = \sum_{j(\neq i)} (w_{ij}p_j - w_{ji} p_i),
\label{17}
\eeq
where $w_{ij}\geq0$ is the rate of the transition
$j\to i$.
Defining $w_{ii}$, absent in (\ref{17}), in such a way that
\beq
\sum_i w_{ij} = 0,
\label{15}
\eeq
then the master equation can be written in the form
\beq
\frac{dp_i}{dt} = \sum_j w_{ij} p_j,
\label{13}
\eeq
where $w_{ij}$ are understood as the elements of a matrix,
the evolution matrix. From equation (\ref{15}), it follows
that the diagonal elements of the evolution matrix are
negative or zero, $w_{ii}\leq0$. We will consider only
transitions that have the reverse. That is, if $w_{ij}$
is nonzero so is $w_{ji}$.

We denote by $q_i$ the stationary solution of the master
equation. It fulfills the equation
\beq
\sum_j w_{ij} q_j = 0.
\label{13a}
\eeq

The transition rates are set up according to several processes
that causes a change in the state of the system. We consider
first the processes associated to the contact of the system
with heat reservoirs at distinct temperatures. If we let $E_i$
be a state function representing the energy of the system
then the transition rate associated to the $k$ reservoir
at a temperature $T_k$ is
\beq
a_{ij}^k = A_{ij}^k e^{-\beta_k (E_i-E_j)/2},
\label{36}
\eeq
where $A_{ji}^k=A_{ij}^k$ and $\beta_k=1/\kappa T_k$,
and $\kappa$ is the Boltzmann constant.
From this relation it follows that the ratio between
the forward and backward transition rates is  
\beq
\frac{a_{ji}^k}{a_{ij}^k} = e^{-\beta_k(E_j-E_i)}.
\label{31}
\eeq

We now consider the transitions associated to external
actions done on the system. To this end we suppose that
the system is acted by an external potential $V_i$
due to external forces $\mu_l$, that is,
\beq
V_i = - \sum_l \mu_l N_i^l,
\eeq
where $N_i^l$ are some state functions representing the
quantity that changes by the action of the force $\mu_l$.
The transition rate associated to the change of $N_i^l$ is
\beq
b_{ij}^l = B_{ij}^l e^{-\beta_l (E_i-E_j)/2 -\beta_l (V_i-V_j)/2},
\label{36b}
\eeq
where $B_{ij}^l=B_{ji}^l$.
We assume that this transition changes only $N_i^l$ whereas
the other variables $N_i^{l'}$, $l'\neq l$, remain unchanged.
That is, $B_{ij}^l$ vanishes whenever $N_i^{l'}\neq N_i^{l'}$
for $l'\neq l$. In view of these restrictions, equation
(\ref{36b}) becomes
\beq
b_{ij}^l = B_{ij}^l e^{-\beta_l (E_i-E_j)/2
+ \beta_l \mu_l(N_i^l-N_j^l)/2}.
\label{37}
\eeq
From this equation, the ratio of
the forward and backward transition rate is
\beq
\frac{b_{ji}^l}{b_{ij}^l} = e^{-\beta_l (E_j-E_i)
+ \beta_l\mu_l(N_j^l-N_i^l)}.
\label{36c}
\eeq

The transition rate $w_{ij}$ is the sum of the transition rates
just introduced,
\beq
w_{ij} = \sum_k a_{ij}^k + \sum_l b_{ij}^l
\label{35}
\eeq
and we point out that, given a transition $j\to i$, then just
one term on the right-hand side of (\ref{35}) can be nonzero.
This assumption is accomplished by the partition of the whole
set of possible transitions $j\to i$ in several disjoint subsets,
each one associated to a given process. In other words, 
given a transition $j\to i$ it is carried out by only one
of the processes.

\subsection{Heat flux}

Let us determine the time derivative of the average
$U=\la E_i\ra$. From the master equation, and using
(\ref{35}), we find
\beq
\frac{dU}{dt} = \Phi_{\rm u} = \sum_k \phi_k + \sum_l \phi_l,
\label{38}
\eeq
where
\beq
\phi_k = \sum_{ij}(E_i-E_j) a_{ij}^k p_j,
\label{51}
\eeq
and
\beq
\phi_l= \sum_{ij}(E_i-E_j) b_{ij}^l p_j.
\label{52}
\eeq
Equation (\ref{38}) is identified with equations
(\ref{39a}) and (\ref{39b}).

Let us now determine the time derivative of the average
$N_l=\la N_i^l\ra$. From the master equation, and using
(\ref{35}), we find
\beq
\frac{dN_l}{dt} = \chi_l + R_l,
\label{44}
\eeq
where
\beq
\chi_l= \sum_{ij}(N_i^l-N_j^l) b_{ij}^l p_j
\label{53}
\eeq
is understood as the flux of $N_l$ into the system, and
\beq
R_l = \sum_k \sum_{ij}(N_i^l-N_j^l) a_{ij}^k p_j
\eeq
is understood as the creation or annihilation of $N_l$
by the internal processes represented by the rates
$a_{ij}^k$. This last formula was obtained using the condition
stated just below equation (\ref{36b}). Equation (\ref{44})
is identified with equation (\ref{25}).

The total work done on the system per unit time is 
\beq
\Phi_{\rm w} = \sum_l \mu_l \chi_l,
\eeq
which is identified as equation (\ref{45}), and the total
heat flux $\Phi=\Phi_{\rm u}-\Phi_{\rm w}$ is
\beq
\Phi = \sum_k \phi_k + \sum_l (\phi_l - \mu_l \chi_l),
\eeq
which is identified with equation (\ref{27}). 

\subsection{Entropy production and entropy flux}

The entropy of the system is assumed to be given by
the Gibbs formula,
\beq
S = - \kappa \sum_i p_i \ln p_i.
\eeq
Its time derivative is
\beq
\frac{dS}{dt}
= - \kappa \sum_{ij} w_{ij} p_j \ln p_i.
\label{26}
\eeq
Using property (\ref{15}), it can be written in the
equivalent form,
\beq
\frac{dS}{dt} = \kappa \sum_{ij} w_{ij}p_j \ln \frac{p_j}{p_i}.
\label{9}
\eeq

The variation of the entropy with time is
split into two parts,
\beq
\frac{dS}{dt} = {\cal P} - \Psi,
\eeq
where ${\cal P}$ is the entropy production rate and $\Psi$
is the entropy flux from the system to the outside. The
entropy production rate is postulated to be given by
the Schnakenberg formula \cite{schnakenberg1976}
\beq
{\cal P} = \frac\kappa2\sum_{ij} (w_{ij}p_j - w_{ji} p_i)
\ln\frac{w_{ij}p_j}{w_{ji}p_i}.
\label{6}
\eeq
We point out that each term of the summation in
(\ref{6}) is nonnegative because it is of the type
$(x-y)\ln x/y\geq0$. Therefore,  ${\cal P}\geq0$,
which justify the postulate of irreversible 
thermodynamics given by equation (\ref{19}).

The production of entropy can also be expressed in the form
\beq
{\cal P} = \kappa\sum_{ij} w_{ij}p_j
\ln\frac{w_{ij}p_j}{w_{ji}p_i}.
\label{11a}
\eeq
From this expression of $\cal P$ and from (\ref{9}), we
obtain the expression for the entropy flux, which is
\beq
\Psi = \kappa\sum_{ij} w_{ij}p_j \ln\frac{w_{ij}}{w_{ji}},
\label{11b}
\eeq
and we see that it holds the important property of being
linear in $p_i$.

Replacing (\ref{35}) in equation (\ref{11b}) we may write the
entropy flux as
\beq
\Psi = \kappa \sum_k \sum_{ij} a_{ij}^k p_j
\ln\frac{a_{ij}^k}{a_{ji}^k}
+ \kappa \sum_l \sum_{ij} b_{ij}^l p_j \ln\frac{b_{ij}^l}{b_{ji}^l}.
\eeq
Using (\ref{31}) and (\ref{36c}), we find
\[
\Psi = - \sum_k\sum_{ij} \frac{a_{ij}^k}{T_k} (E_i-E_j)p_j
\]
\beq
- \sum_l \sum_{ij} \frac{b_{ij}^l}{T_l} (E_i-E_j) p_j
+ \sum_l \sum_{ij} \frac{b_{ij}^l}{T_l} \mu_l (N_i^l-N_j^l)p_j,
\eeq
where we used again the condition stated just below equation
(\ref{36b}), and $T_l=1/\kappa\beta_l$. Using (\ref{51}),
(\ref{52}), and (\ref{53}), this equation can be written as
\beq
\Psi = - \sum_k \frac{1}{T_k}\phi_k
- \sum_l \frac{1}{T_l}  (\phi_k - \mu_l \chi_l),
\label{21b}
\eeq
which is identified with equation (\ref{21}). Therefore, we
may say that the Clausius relation introduced in \ref{CL} as a postulate
of the present formulation of irreversible thermodynamics in order
to reach equation (\ref{21}) is a direct consequence of the
form we have assumed for the transitions rates, namely that
given by equations (\ref{36}) and (\ref{36b}).

\subsection{Convexity of ${\cal P}$ in relation to $p_i$}

We show now that the production of entropy is an
upward convex function of the collection of $p_i$.
That is, we show that the second order variation 
of the entropy production in relation to variations
in $p_i$ is nonnegative, $\delta^2{\cal P}\geq 0$.
To this end we first observe that $\Psi$ is linear
in $p_i$ from which follows that its second order
variation in relation to variations in
$p_i$ vanishes. Therefore
\beq
\delta^2{\cal P} = \delta^2 \Gamma,
\eeq
where $\Gamma=dS/dt={\cal P}-\Psi$ is the expression
on the right-hand side of (\ref{26}), that is,
\beq
\Gamma = - \kappa\sum_{ij} w_{ij} p_j \ln p_i.
\label{16}
\eeq
The second order variation of $\Gamma$ is
\beq
\delta^2 \Gamma = \frac\kappa2 \sum_{ij} \frac{\partial^2\Gamma}
{\partial p_i\partial p_j}\delta p_i\delta p_j.
\label{18}
\eeq

From (\ref{16}), we obtain
\beq
\frac{\partial\Gamma}{\partial p_j} = - \kappa\sum_{k} w_{kj} \ln p_k
- \kappa \sum_{k} w_{jk}  \frac{p_k}{p_j},
\eeq
and
\beq
\frac{\partial^2\Gamma}{\partial p_i\partial p_j}
= -  \kappa(\frac{w_{ij}}{p_i} + \frac{w_{ji}}{p_j})
+ \kappa\,\delta_{ij} \sum_{k} w_{jk}  \frac{p_k}{p_j^2}.
\label{54}
\eeq
Replacing this result in (\ref{18}), we find
\beq
\delta^2 {\cal P} = \delta^2 \Gamma
= \frac\kappa2 \sum_{ij} w_{ij} p_j
(\frac{\delta p_i}{p_i} - \frac{\delta p_j}{p_j})^2,
\eeq
where we used the property (\ref{15}). Taking into account
that $w_{ij}\geq0$ for $i\neq j$, we reach the desired result
\beq
\delta^2{\cal P} = \delta^2\Gamma \geq 0.
\label{40}
\eeq

\subsection{Convexity of ${\cal P}$ in relation to the fluxes}

We have just proven that ${\cal P}$ is convex in relation to
the probabilities $p_i$. We now show that ${\cal P}$ is also
convex in relation to the fluxes $J_k$. This property is
expected because the fluxes are linear in $p_i$.

As before, we use the notation $J_k$ for $-\phi_k$ or
$-\chi_l$ and the notation $F_k$ for $1/T_k$ or $-\mu_k/T_k$,
already introduced above. Then the expression (\ref{21b}) for
$\Psi$ is written in the simplified form
\beq
\Psi = \sum_k F_k J_k.
\eeq
As the fluxes $J_k$ are linear in $p_j$,
it can be written as 
\beq
J_k = \sum_j  f_{kj} p_j,
\eeq
and the explicit expressions for the coefficients $f_{kj}$
are obtained from (\ref{51}), (\ref{52}), and (\ref{53})
and they are either
\beq
f_{kj} = - \sum_i (E_i-E_j) a_{ij}^k,
\eeq
or
\beq
f_{lj} = - \sum_i(E_i-E_j) b_{ij}^l,
\eeq
or
\beq
f_{lj} = - \sum_i(N_i^l-N_j^l) b_{ij}^l.
\eeq

Let us define as before 
\beq
X_k = \frac{\partial{\cal P}}{\partial J_k},
\eeq
and
\beq
A_{kl} =  \frac{\partial^2{\cal P}}{\partial J_k\partial J_l}
=  \frac{\partial X_k}{\partial J_l}.
\eeq

We also define 
\beq
D_{ij} = \frac{\partial^2{\cal P}}{\partial p_i\partial p_j}
= \frac{\partial^2\Gamma}{\partial p_i\partial p_j},
\eeq
the explicit form of which is given by (\ref{54}),
from which we obtain 
\beq
\delta^2 {\cal P} = \frac12 \sum_{ij} D_{ij} \delta p_i \delta p_j.
\label{55}
\eeq
The relation between $D_{ij}$ and $A_{kl}$ is
\beq
D_{ij} = \sum_{kl} A_{kl} f_{ki} f_{lj},
\eeq
which follows because $J_k$ is linear in $p_i$.
Replacing this relation in (\ref{55}), we find
\beq
\delta^2 {\cal P} = \frac12 \sum_{kl} A_{kl} \delta J_k \delta J_l,
\eeq
where
\beq
\delta J_k = \sum_i  f_{ki} \delta p_i.
\eeq
But $\delta^2{\cal P}\geq0$, as shown above in (\ref{40}), and
\beq
\sum_{kl} A_{kl} \delta J_k \delta J_l \geq 0,
\eeq
which proves that ${\cal P}$ is an upward convex function of
the collection of variables $J_k$, which we have taken as a
postulate of the present approach to irreversible thermodynamics. 
taken as an assumption just above equation (\ref{10}).
From this property, follows the Glansdorff-Prigogine principle
shown in \ref{GP}.

\subsection{Excess entropy production}

Let us define the quantity $C_i$ by
\beq
C_j = \frac{\partial{\cal P}}{\partial p_j}.
\label{56}
\eeq
From the definition of ${\cal P}$, given by (\ref{6}), we find
\beq
C_j = \kappa\sum_i w_{ij} \ln\frac{w_{ij}p_j}{w_{ji}p_i}
- \kappa\sum_i  w_{ji}\frac{p_i}{p_j},
\eeq
where we used the property (\ref{15}). In the 
stationary state, $p_i=q_i$, the value of $C_j$ is
\beq
C_j^0 = \kappa\sum_i w_{ij} \ln\frac{w_{ij}q_j}{w_{ji}q_i},
\label{41}
\eeq
where we have used property (\ref{13a}).

The excess entropy production is defined by
\beq
{\cal P}_{\rm exc} = {\cal P} - \sum_j C_j^0(p_j-q_j) - {\cal P}_0,
\label{42}
\eeq
where ${\cal P}_0$ is the stationary value of ${\cal P}$,
\beq
{\cal P}_0 = \kappa\sum_{ij} w_{ij}q_j \ln\frac{w_{ij}q_j}{w_{ji}q_i}.
\label{43}
\eeq
Taking into account that ${\cal P}_{\rm exc}$ differs from 
${\cal P}$ by linear terms then
\beq
\delta^2{\cal P}_{\rm exc}=\delta^2{\cal P},
\eeq
and it is also an upward convex function of $p_i$. Taking into
account that ${\cal P}_{\rm exc}$ vanishes at the stationary
state and that its variation at the stationary state also
vanishes, $\delta {\cal P}_{\rm exc}=0$, then we may write
\beq
{\cal P}_{\rm exc} \geq 0.
\eeq

Using the relation
\beq
\frac{\partial{\cal P}}{\partial p_j} = \sum_k
\frac{\partial{\cal P}}{\partial J_k}
\frac{\partial J_k}{\partial p_j},
\eeq
we find
\beq
C_j =\sum_k X_k f_{kj},
\eeq
from which we get
\beq
\sum_j C_j^0(p_j-q_j) = \sum_k X_k^0 (J_k - J_k^0),
\eeq
which replaced in (\ref{42}) gives
\beq
{\cal P}_{\rm exc} = {\cal P}
- \sum_k X_k^0 (J_k - J_k^0) - {\cal P}_0,
\eeq
which is identified as the expression (\ref{28}),
and the excess entropy defined within our formulation of 
irreversible thermodynamics coincides with the
excess entropy defined by expression (\ref{42}).

\subsection{Lyapunov function}

Replacing in equation (\ref{42}) the expression for $C_j^0$,
given by (\ref{41}), and the expressions for ${\cal P}$ and
${\cal P}_0$, given by (\ref{11a}) and (\ref{43}), we find
\beq
{\cal P}_{\rm exc} = - \kappa\sum_{ij} w_{ij}p_j \ln\frac{p_i}{q_i},
\eeq
where we used the property (\ref{13a}).
Using the master equation in the form (\ref{13}), we see that
the right-hand side of this equation is the time derivative of
\beq
L = - \kappa \sum_i p_i\ln \frac{p_i}{q_i},
\eeq
that is,
\beq
{\cal P}_{\rm exc} = \frac{dL}{dt},
\label{59}
\eeq
from which follows
\beq
\frac{dL}{dt}\geq0,
\label{46a}
\eeq
because ${\cal P}_{\rm exc}\geq0$.

The quantity $L$ can yet be written in the form
\beq
L = - \kappa \sum_i [p_i \ln \frac{p_i}{q_i} - (p_i - q_i)],
\eeq
from which follows that
\beq
L\leq0,
\label{46b}
\eeq
because the expression inside
square brackets is greater or equal zero.
The two inequalities (\ref{46a}) and (\ref{46b}) show
that $L$ is a Lyapunov function in relation to
the variables $p_i$.

\section{Absence of temperature grandients}

The equation (\ref{59}) tell us that the excess entropy
production is a time derivative of $L$ which is a function
of the probabilities $p_i$. We may ask whether
it is possible to write the excess entropy production
as a time derivative of a thermodynamic potential, understood
as a linear combination of $S$, $U$ and $N_l$. This is indeed
possible if the processes are isothermal, that is, if the
heat introduced into the system comes from reservoirs that
have all the same temperature which means that the system
is subject to no gradients of temperature. In other words,
$\beta_k$ and $\beta_l$ appearing in the rates (\ref{36})
and (\ref{37}) should have the same value.

When $\beta_k=\beta$, independent of $k$, the several 
transitions given by (\ref{36}) can be gathered into
a single transition rate $a_{ij}$, given by
\beq
a_{ij} = A_{ij} e^{-\beta (E_i-E_j)/2},
\label{48}
\eeq
where $A_{ij}=A_{ji}$. Considering that $\beta_l=\beta$
is also independent of $l$, the transitions (\ref{37})
becomes
\beq
b_{ij}^l = B_{ij}^l e^{-\beta (E_i-E_j)/2
+ \beta \mu_l(N_i^l-N_j^l)/2},
\label{49}
\eeq
where $\beta=1/\kappa T$, and $T$ is the common temperature
of the reservoirs.

The ratio of the rates of the forward and backward transitions are
\beq
\frac{a_{ij}}{a_{ji}} = e^{-\beta (E_i-E_j)},
\label{48a}
\eeq
\beq
\frac{b_{ij}^l}{b_{ji}^l} = e^{-\beta (E_i-E_j)
+ \beta \mu_l(N_i^l-N_j^l)},
\label{49a}
\eeq
and we remark that these relations are not the condition of 
detailed balance, which means that the stationary state may
be a nonequilibrium stationary state, although the temperatures
of the reservoirs are all the same. The detailed balance
condition is satisfied if the transitions determined by the
rate $a_{ij}$ connects states $i$ and $j$ such that the
external potential are equal, $V_i=V_j$. In this case we
see that both ratios (\ref{48a}) and (\ref{49a})
are the same and given by $p^{\rm e}_i/p^{\rm e}_j$ where
$p_i^{\rm e}$ is proportional to $e^{-\beta (E_i + V_i)}$
and understood as the equilibrium probability distribution.
In the case of a chemical system, the condition $V_i=V_j$
for internal processes leads to the well known relation
between the chemical potentials of and the stoichiometric
coefficientsexpressing the equilibrium condition
\cite{tome2018}.

In the present case, the total transition rate is written as
\beq
w_{ij} = a_{ij} + \sum_l b_{ij}^l,
\eeq
and again, given $i$ and $j$ only one term on the right-hand
side of this equation can be nonzero. The time variation of
the energy is
\beq
\frac{dU}{dt} = \Phi_{\rm u},
\label{50}
\eeq
where
\beq
\Phi_{\rm u} = \sum_{ij}(E_i-E_j)w_{ij}.
\eeq
The flux of heat is
\beq
\Phi = \Phi_{\rm u} - \sum_l \mu_l \chi_l.
\eeq

From the formula (\ref{11b}) for the entropy flux 
and using (\ref{48a}) and (\ref{49a}), we obtain 
\[
\Psi = - \frac1T \sum_{ij} w_{ij}p_j (E_i-E_j)
\]
\beq
+  \frac1T \sum_l \mu_l \sum_{ij} b_{ij}^l p_j (N_i^l-N_j^l)],
\eeq
which can be written as
\beq
\Psi = - \frac1T \Phi_{\rm u} +  \frac1T \sum_l \mu_l \chi_l.
\eeq

Before we proceed to determine other quantities of interest,
we observe that the entropy flux at the stationary state is
\beq
\Psi_0 = - \frac1T \Phi_{\rm u}^0 + \frac1T \sum_l\mu_l\chi_l^0.
\eeq
Subtracting these two equations, we find
\beq
\Psi-\Psi_0 = - \frac1T (\Phi_{\rm u} -\Phi_{\rm u}^0)
+ \frac1T \sum_l\mu_l(\chi_l-\chi_l^0).
\label{57}
\eeq

The excess entropy production ${\cal P}_{\rm exc}$ is given by
(\ref{42}), and is
\beq
{\cal P}_{\rm exc} = {\cal P} - \sum_j C_j^0(p_j-q_j) - {\cal P}_0,
\eeq
which we write as
\beq
{\cal P}_{\rm exc} = \frac{dS}{dt} + \Psi
- \sum_j C_j^0(p_j-q_j) - \Psi_0,
\eeq
because ${\cal P} = dS/dt+ \Psi$ and ${\cal P}_0=\Psi_0$.
Now from the definition of $C_j$ given by (\ref{56})
\beq
C_j = x g_j + \sum_l y_l h_j^l,
\eeq
where
\beq
x=\frac{\partial{\cal P}}{\partial \Phi_{\rm u}},
\qquad
y_l=\frac{\partial{\cal P}}{\partial \chi_l},
\eeq
and
\beq
g_j = \frac{\partial\Phi_{\rm u}}{\partial p_j},
\qquad
h_j^l = \frac{\partial\chi_l}{\partial p_j}.
\eeq

Considering that $\Phi_{\rm u}$ and $\chi_l$ are linear
in $p_i$ then the coefficients $g_j$ and $h_j^l$ are
independent of $p_i$. This property allow us to 
write
\beq
C_j^0 = x^0 g_j + \sum_l y_l^0 h_j^l,
\eeq
where $x^0$ and $y_l^0$ are the values of $x$ and $y_l$
at the stationary state, that is, when $p_i=q_i$.
Using the linear property we may also write
\beq
\Phi_{\rm u} = \sum_j g_j p_j,
\eeq
\beq
\chi_l = \sum_j h_j^l p_j,
\eeq
which lead us to the following conclusion
\beq
\sum_j C_j^0(p_j-q_j) = x^0 (\Phi_{\rm u} -\Phi_{\rm u}^0)
+ \sum_l y_l^0 (\chi_l-\chi_l^0).
\eeq

From this last result and from the expression (\ref{57})
for $\Psi-\Psi_0$ we reach the following expression
for the excess entropy production
\beq
{\cal P}_{\rm exc} = \frac{dS}{dt}
- \frac{r}T (\Phi_{\rm u} -\Phi_{\rm u}^0)
+ \sum_l \frac{\alpha_l}{T}(\chi_l-\chi_l^0),
\eeq
where we are using the abbreviations
\beq
\frac{r}T = \frac1T + x^0 \qquad \frac{\alpha_l}{T}
= \frac{\mu_l}{T} - y_l^0.
\eeq

In view of equations (\ref{50}) and (\ref{44}),
it can be written as
\beq
{\cal P}_{\rm exc} = \frac{dS}{dt}
- \frac{r}T \frac{dU}{dt}
+ \sum_l \frac{\alpha_l}{T}\frac{dN_l}{dt},
\eeq
bearing in mind that $\Phi_{\rm u}^0=0$ and that
$R_l = -\chi_l^0$.

This last equality allows us to write 
\beq
{\cal P}_{\rm exc} = \frac{dM}{dt},
\label{58}
\eeq
where
\beq
M = S - \frac{r}T U + \sum_l \frac{\alpha_l}{T} N_l +K,
\label{60} 
\eeq
where $K$ is a constant.
That is, the excess entropy production is the
time derivative of $M$ which is a linear combination
of $S$ and $U$, and the complementary variables $N_l$,
and can then be understood as a thermodynamic potential.

Comparing equations (\ref{58}) and (\ref{59}), we see
that $M$ and $L$ differ by a constant. Since $L$
vanishes at the stationary state, we conclude that
$L=M-M_0$ where $M_0$ is the value of $M$ at the
stationary state. Since $L\geq0$ then
\beq
M \geq M_0,
\eeq
and $M-M_0$ can be understood
as a Lyapunov function in relation to the fluxes
because $S$, $U$, and $N_l$ are functions of the
fluxes.

\section{Conclusion}

We have derived the main equations of irreversible thermodynamics
from stochastic thermodynamics including the Glansdorff-Prigogine
extremal principle. To this end we used a Master equation 
defined through transition rates that represent the various
processes that are induced by gradients of temperature and
external forces. The production of entropy was shown to be
an upward convex function of the probabilities of the microstates.
Considering that the fluxes are linear in these
probabilities we showed that the entropy production can
also be understood as a convex function of the fluxes.
The convexity property is then used to show that
the excess entropy production is a minimum at the stationary
state, which is a statement of the Glansdorff-Prigogine
principle.

The stability of the stationary state can be analyzed by
thinking of the master equation as a set of ordinary
differential equations in the variables $p_i$ and by
the construction of a Lyapunov function in the variables $p_i$.
We have construct such an equation and showed that
its time derivative is equal to the excess entropy production.
The question we have raised is whether we can construct
a Lyapunov function in relation to the macroscopic variables
$S$, $U$ and $N_l$, that is, a Lyapunov function associated
to the set of ordinary differential equations in these variables. 
We have shown that this is possible when the temperatures
associated to the various transitions
rates are the same, that is, when no gradients of temperature
are present. In this case the Lyapunov function is a thermodynamic
potential in the sense that it is a linear combination of
$S$, $U$ and $N_l$.


\end{document}